\documentclass[12pt,reqno,a4paper]{article}
\usepackage{authblk}
\usepackage[margin=1in]{geometry}
\RequirePackage{amsthm,amsmath,amsfonts,amssymb}
\usepackage{array}
\RequirePackage{natbib}
\RequirePackage[colorlinks,citecolor=blue,urlcolor=blue]{hyperref}
\usepackage{booktabs,longtable}
\usepackage{graphicx}
\usepackage{xcolor}
\usepackage{epstopdf}
\usepackage[inline]{enumitem}
\usepackage{bm}
\usepackage[ruled, lined]{algorithm2e}
\usepackage{setspace}
\usepackage{subcaption}

\graphicspath{{./}{../figures/}}

\allowdisplaybreaks

\newcommand{\vA}{\bm{A}}
\newcommand{\vB}{\bm{B}}
\newcommand{\vtheta}{\bm{\theta}}
\newcommand{\valpha}{\bm{\alpha}}
\newcommand{\Prob}{\mathrm{\bf Pr}}
\newcommand{\vc}{\bm{c}}
\newcommand{\vx}{\bm{x}}
\newcommand{\vy}{\bm{y}}
\newcommand{\vZ}{\bm{Z}}
\newcommand{\ER}{Erd\"{o}s-R\'{e}nyi}

\newcommand{\given}{\, | \,}
\DeclareMathOperator*{\argmax}{argmax}

\title{A Mixed-Membership Model for Social 
Network Clustering}

\author[1]{Guang Ouyang}
\author[1]{Dipak K.\ Dey}
\affil[1]{Department of Statistics, University of Connecticut, 
Storrs, CT 06269, USA}
\author[2]{Panpan Zhang\thanks{Corresponding author. Email: 
panpan.zhang@vumc.org}}
\affil[2]{Department of Biostatistics, Vanderbilt University Medical 
Center, Nashville, TN 37203, USA}

\begin{document}

\maketitle

\begin{abstract}
  We propose a simple mixed membership model for social network 
  clustering in this paper. A flexible function is adopted to 
  measure affinities among a set of entities in a social network. 
  The model not only allows each entity in the network to possess 
  more than one membership, but also provides accurate statistical 
  inference about network structure. We estimate the membership 
  parameters using an MCMC algorithm. We evaluate the performance of 
  the proposed algorithm by applying our model to two empirical 
  social network data, the Zachary club data and the bottlenose 
  dolphin network data. We also conduct some numerical studies based 
  on synthetic networks for furtherassessing the effectiveness of 
  our algorithm. In the end, some concluding remarks and future work 
  are addressed briefly.
\end{abstract}


\section{Introduction}
\label{Sec:Intro}
Social network analysis is part of the social science which 
is an academic discipline studying a society and the behavior of 
entities therein. A social network consists of a set of entities 
(called actors) with certain interactions (represented by 
ties) among them. Statistical modeling has been a popular and 
powerful tool to study social networks thanks to its solid 
theoretical foundation. A plethora of statistical models have been 
established and exploited to uncover relational structure of social 
networks, and dyadic ties among actors. Friendship among Facebook 
users, business relationship across companies on the Wall 
street, and collaborations among researchers in a scientific field 
are all social network examples that have been extensively studied 
in the past. Social network analysis has a long history in 
sociology, where classical works traced back to the 1940s and 
1950s~\citep{Rapoport49a, Rapoport49b, Rapoport50, Harary, 
Cartwright}.

Modern research on social network analysis 
within mathematics, physics and other scientific disciplines focus mainly
on the following three distinctive network features. The first 
feature is to explore how local mechanisms of network formation 
produce global network structure. Two representative models are the 
network evolution model~\citep{Newman98} and the 
nodal attribute model~\citep{Boguna}. We refer the readers 
to~\citet{Toivonen} for a comparison of these two models, and to 
the survey paper~\citep{Snijdersrev} for a complete review of 
related statistical models. 
The second feature is to investigate topological properties of 
social networks and develop methods of modeling, either analytically 
or numerically. Two of the most popular properties of social 
networks are the small world phenomenon~\citep{Watts} and the 
power law of degree distribution~\citep{Barabasi}. A summary 
of some solvable random-graph-based social network models was given 
in~\citep{Newman02}. 
The third feature, which is the one that we investigate in 
this paper, is network clustering.

Social network clustering works under the rationale that a group of
actors excessively tied in a network are inclined to forming a cluster.
One of the seminal works on social network clustering
was~\citet{Watts}, where each pair of actors in a social network
was proven to be tied with a high probability if they had a mutual 
acquaintance, and such ``tie'' was parameterized by a measure 
called the clustering coefficient. This natural phenomenon in 
social networks was also discussed extensively 
by~\citet{Newman98,Newman01}. The formation of a cluster requires the 
connections of actors within the cluster are significantly higher 
than those between actors from different clusters. It was posited 
in some literatures, e.g.,~\citep{Holland}, that a high probability 
of the occurrence of ties between actors within a cluster was due to 
some kind of homology (also called ``internal homogeneity'') 
of the actors. For instance, students from the same department of a 
college tend to form a community, in which almost everybody is a 
friend of everybody (i.e., the students in the same community are 
more likely to be connected); while students with different 
educational background are much less likely to be connected. Such 
internal homogeneity is mostly reflected in a background parameter 
(e.g., same department) and a location parameter (e.g., same 
college).

In this paper, we propose a simple but effective method for 
accurately clustering the entities in a social network into mutually 
exclusive communities. The proposed model was inspired and elevated 
from the classical stochastic blockmodel~\citep[SBM,][]{Nowicki}. 
Recently, there were a variety of 
models extended from SBM in the literature. For instance, 
\citet{Sengupta2018ablock} introduced an SBM adjusted by node 
popularity, \citet{Huang2020mixed} established an SBM for 
heterogeneous networks accounting for node attribute and 
\citet{Noroozi2022thehierarchy} suggested a nested 
SBM integrating standard SBM and LSM. Different from the existing 
literature, we specifically consider a 
flexible function to measure the similarities between actors in a 
network. Mixed membership is allowed for each actor in our model. 
The fit of our model is done in a Bayesian framework. The ascendancy 
of our model over the classic SBM will be detailed and discussed in 
the subsequent section. This paper not only introduces a flexible 
and 
extensible model allowing mixed memberships for network actors, but 
also gives the interested researchers, especially those relatively 
new to the field, insights into a standard approach of conducting 
statistical inference for social network clustering problems.

The rest of this paper is organized as follows: We review some 
representative model-based methods for social network 
clustering, with an additional concentration on the SBM, in 
Section~\ref{Sec:review}. We propose a mixed membership model 
based on a simple similarity function in Section~\ref{Sec:distance}. 
Theoretical parameter estimation and an associated MCMC 
algorithm are presented in Section~\ref{Sec:estimate}. Two empirical 
social network examples, the Zachary karate club data and the 
bottlenose dolphin network data, are used to evaluate the 
performance of our model, shown respectively in 
Sections~\ref{Sec:zachary} and~\ref{Sec:dolphin}. We then 
conduct some simulation study on synthetic data in 
Section~\ref{Sec:simulation}. In the end, we give some concluding 
remarks and propose some future work in Section~\ref{Sec:concluding}.

\section{Notations for Stochastic Blockmodels}
\label{Sec:review}
In general, methods for social network clustering can be 
summarized into two categories. A metric-based 
method, in contrast, aims at specifying an objective function 
which evaluates the quality of each network clustering strategy, 
followed by an algorithm optimizing the objective function
\citep[e.g.,][]{Ng, Shi, Newman, Ouyang}.
A model-based method is to 
propose a (parametric) graphical generative model that characterizes 
the community structure of a social network, followed by an
algorithm estimating the membership parameters conditioning on the 
observed data, most done in a Bayesian framework. To date, there 
have emerged a variety of graphical models for social network 
clustering, including but not limited to 
stochastic blockmodels~\citep[SBM,][]{Nowicki, Airoldi, 
	Abbe2018community, Gao2018community}, latent 
	space models~\citep[LSM,][]{Hoff2002latent, Handcock, 
	Sewell2017latent} random dot product 
graphs~\citep[RDPG,][]{Young2007random, Marchette2008predicting, 
	Lyzinski2017community, Athreya2018statistical}, and exponential 
random graph models~\citep[ERGM,][]{Snijders2006new, Hunter2008ergm,
	Fronczak2013exponential}, among others.

The core idea model-based methods
is to theoretically uncover the probabilistic and 
statistical properties of the proposed models. To begin with, we 
introduce some notations that will be used all through the paper. 
In general, a network is modeled by a mathematical undirected 
(or directed) graph consisting of a set of nodes which represent 
actors (e.g., Facebook users) in the network, and a set of 
undirected (or directed) edges which represent the relational ties 
between each pair of nodes (e.g., friendship connections between 
Facebook users). Let $n$ be the number of nodes in an undirected 
social network. The observation of the network can be mathematically 
represented by an $n \times n$ dyadic adjacency matrix
$\vA = (A_{ij})_{n \times n}$, where $A_{ij}$ equals $1$ if nodes 
$i$ and $j$ are connected; $0$, otherwise. For undirected networks, 
adjacency matrices are symmetric. If a network is directed, 
$A_{ij} = 1$ refers to a directed relation from $i$ (initiator) to 
$j$ (receiver), and the associated adjacency matrix $\vA$ may be 
asymmetric.

More specifically, we consider the stochastic blockmodel, first 
proposed by~\citet{Snijders}. Although directed networks were 
considered in~\citet{Snijders}, we simplify the problem to 
undirected networks for the sake of explanation. The model and the 
related methods can be extended to directed networks effortlessly. 
Our goal is to cluster a network of order $n$ into $h$ distinct 
communities. For each node $i = 1, 2, \ldots, n$, let $c(i)$ 
denote the community membership function for $i$. Assuming that 
$c(1), c(2), \ldots, c(n)$ are independently and identically distributed
(i.i.d.) multinomial random variables
with a hyperparameter vector
$\vtheta = (\theta_1, \theta_2, \ldots, \theta_h)$,
one defines $\vB$ as an $h \times h$ symmetric 
probability matrix indicating linkages across different 
communities. Conditioning on $c(1), c(2), \ldots, c(n)$, the 
distribution of $A_{ij}$ for each pair of nodes $i$ and $j$ is 
Bernoulli with probability $B_{c(i)c(j)}$.

As $\vA$ is 
observed, our goal turns to estimate hyperparameters in $\vtheta$ 
and the probability matrix $\vB$, and ultimately to uncover the 
network structure by inferring $\vc = (c(1), c(2), \ldots, c(n))$. 
The estimation can be performed in a Bayesian framework:
\begin{enumerate}
	\item Establish the likelihood function, $\Prob(\vA, \vc; 
	\vtheta, \vB)$;
	\item Estimate $\vtheta$ and $\vB$ jointly, which usually 
	can be done by some Bayesian methods, such as Markov Chain Monte 
	Carlo (MCMC) algorithms;
	\item Determine the posterior distribution of $\vc$ given 
	$\vA$, which is given by
	\[\Prob(\vc \given \vA) \propto \int \Prob(\vA, \vc \given 
	\vtheta, \vB) \pi(\vtheta, \vB) \, d\vtheta d\vB,\]
	where $\pi(\vtheta, \vB)$ denotes a joint prior distribution 
	of $\vtheta$ and $\vB$.
\end{enumerate}
The community prediction of node $i$ is the index of the 
membership with the largest posterior probability.

There are several shortcomings of the classical SBM. One is that 
each actor in the network only can be assigned to one community, 
which may not be the case for many real social networks. A mixed 
membership SBM, inspired from the latent Dirichlet 
allocation~\citep[LDA,][]{Blei}, was proposed by~\citet{Airoldi} to 
break this limitation. The model in~\citet{Airoldi} allows each 
actor in the network to possess multiple community memberships. In 
addition, it seems natural to define a function to quantitatively 
measure the similarities (or dissimilarities) between the actors in 
a social network space. Such functions are viewed as an 
indispensable part in clustering analysis, but are not considered in 
the classic SBM. In Section~\ref{Sec:distance}, we propose a mixed 
membership probabilistic model based on a simple and well-defined 
similarity function.

\section{A Mixed Membership Model}
\label{Sec:distance}
In this section, we propose a simple generative model which 
admits multiple membership (of actors) for social network 
clustering. The development of the model is based on a probabilistic 
relationship between the observed adjacency matrix $\vA$ and a 
similarity function---more specifically, the cosine similarity.

We start by introducing some additional notations and 
preliminaries. Consider a social network consisting of $n$ nodes to 
be clustered into $h$ distinct communities with $h \le n$. For each 
node $i = 1, 2, \ldots, n$, let
$Z_i = (Z_{i1}, Z_{i2}, \ldots, Z_{ih})^{\top}$, where $\sum_{k = 
1}^{h} Z_{ik} = 1$, be an $h \times 1$
vector that represents the mixed membership of node~$i$ across $h$
communities. To be specific, for $1 \le k \le h$,
$Z_{ik}$ refers to the probability that node $i$ belongs to 
community
$k$. The very special case $Z_{ik} = 1$ for some $k$ indicates that 
node $i$ is assigned to community $k$ with probability $1$ without 
uncertainty, though it is very rare in practice.

For any two $s$-dimensional vectors $\vx$ and $\vy$, the cosine 
similarity between $\vx$ and $\vy$ is
\begin{equation}
	\label{Eq:cos}
	\cos(\vx, \vy) = \frac{\vx^{\top} \vy}{||\vx||_2 ||\vy||_2} 
	= \frac{\sum_{r = 1}^{s} x_r y_r}{\sqrt{\sum_{r = 1}^{s} 
			x_r^2}\sqrt{\sum_{r = 1}^{s} y_r^2}},
\end{equation}
where $\|\cdot\|_2$ refers to the standard $\ell_2$ norm. Thus, 
the corresponding dissimilarity function is
$(1 - \mbox{{\rm cosine similarity}})$.

We choose the cosine similarity as the measure of similarity in 
our model for three major reasons:
\begin{enumerate}
	\item Cosine similarity is a simple measure, and it can 
	be easily applied to high-dimensional data.  
	\item Cosine similarity has a standard statistical 
	interpretation, as it is equivalent to the Pearson 
	correlation coefficient for the data that are centered 
	by mean.
	\item Cosine similarity is defined on $[0, 1]$, so it is 
	ready for modeling link density.
\end{enumerate}

Recall the adjacency matrix $\vA = (A_{ij})$. Assuming that 
$A_{ij}$'s are mutually independent, we incorporate a Bernoulli 
model into the link distribution of $A_{ij}$ for nodes $i$ and 
$j$, given their mixed community membership $Z_i$ and $Z_j$; that is,
\begin{equation*}
	\Prob(A_{ij} = a_{ij}) = p(Z_i, Z_j)^{a_{ij}} \left(1 - p(Z_i, 
	Z_j)\right)^{1 - a_{ij}},
\end{equation*}
where
\[p(Z_i, Z_j) = \cos (Z_i, Z_j) = 
\left(\frac{Z_i^{\top}Z_j}{\|Z_i\|_2\|Z_j\|_2}\right)\]
and
\[a_{ij} = 
\begin{cases}
	1, &\mbox{if nodes $i$ and $j$ are connected};
	\\ 0, &\mbox{otherwise}.
\end{cases}\]

By the assumption of conditional independence, we obtain the 
likelihood function of the adjacency matrix $\vA$,
\begin{equation}
	\label{Eq:likelihood}
	\Prob(\vA \given \vZ) = \prod_{1 \le i < j \le n} p(Z_i, 
	Z_j)^{a_{ij}} (1 - p(Z_i, Z_j))^{1 - a_{ij}},
\end{equation}
where $\vZ = (Z_1, Z_2, \ldots, Z_n)$ is an $h \times n$ matrix 
which represents community memberships of all nodes. It is worth 
mentioning that $\vZ$ directly reflects node memberships, so should 
not be interpreted as latent positions for 
LSM~\citep{Hoff2002latent}. Membership parameter and latent position 
are conceptually nonequivalent, though the latter usually has impact 
on network connectivity and is implicitly related to node 
membership. Our goal is to predict $\vZ$ given the observation of 
$\vA$, which can be done via an algorithm shown in 
Section~\ref{Sec:estimate}.

\section{Parameter Estimation}
\label{Sec:estimate}
In this section, we estimate the parameters in our mixed 
membership model via a standard Bayesian method. At first, we posit 
a prior distribution for $\vZ$. Notice that each component in $\vZ$, 
$Z_i$, consists of $h$ elements representing probabilities adding up 
to $1$. Dirichlet distribution appears a reasonable and 
widely-accepted choice for its prior. For $1 \le i \le n$, let 
$Z_i$'s be random variables such that
\[Z_i \overset{i.i.d.}{\sim} {\rm Dirichlet}(\valpha),\]
where $\valpha$ is an $h$-dimensional hyperparameter vector. The 
initial selection of $\valpha$ is flexible unless related 
information is available. In practice, one may choose each element 
in $\valpha$ to be equal to $1/h$. For 
each $Z_i$ in $\vZ$, our goal is to approximate the posterior 
distribution of $Z_i$ given $\vA$. We exploit the Gibbs 
sampling algorithm proposed by~\citet{Gelfand}.

The Gibbs sampling is a well-developed MCMC algorithm, which is 
popular for its simplicity and versatility. The Gibbs sampling 
was first appeared in~\citet{Geman}, and the theoretical properties 
of the algorithm were discussed extensively by~\citet{Casella, 
Gelfand}. It was proven in~\citet{Geman} that the distribution of 
simulated samples converges to the posterior distribution of true 
parameters given the observations, regardless of the starting state 
(i.e., the prior distribution). The key of Gibbs sampling is to 
simulate the next generation of unknown parameters based on the 
estimates at the current state. Let $\vZ^{(m)} = (Z_1^{(m)}, 
Z_2^{(m)}, \ldots, Z_n^{(m)})$ be the estimate in the current 
iteration. We simulate $\vZ^{(m + 1)}$ in the following way:
\begin{enumerate}
	\item Simulate $Z_1^{(m + 1)}$ from the posterior 
	distribution of $Z_1$ given $Z_2^{(m)}, \ldots, Z_n^{(m)}$, 
	$\valpha$ and $\vA$.
	\item For $i = 2, 3, \ldots, n - 1,$ simulate $Z_i^{(m + 1)}$ 
	from the posterior distribution of $Z_i$ given $Z_1^{(m + 1)}, 
	\ldots, \\ Z_{i - 1}^{(m + 1)}$, $Z_{i + 1}^{(m)}, \ldots, 
	Z_n^{(m)}$, $\valpha$ and $\vA$.
	\item Simulate $Z_n^{(m + 1)}$ from the posterior 
	distribution of $Z_1$ given $Z_1^{(m + 1)}, \ldots, Z_{n - 
	1}^{(m + 1)}$, $\valpha$ and $\vA$.
\end{enumerate}

In order to implement the algorithm, we derive the posterior 
distribution of $Z_i$, given $Z_1, \ldots, \\ Z_{i - 1}, Z_{i + 1}, 
\ldots, Z_n$, $\valpha$ and $\vA$. For brevity, denote $Z_{-i} = 
(Z_1, \ldots, Z_{i - 1}, Z_{i + 1}, \ldots, Z_n)$. According to 
the definition of conditional probability, we have
\begin{align}
	\label{Eq:comlikelihood}
	\Prob(Z_i \given Z_{-i}, \valpha, \vA) &= \frac{\Prob(\vZ, \vA 
	\given \valpha)}{\Prob(Z_{-i}, \vA \given \valpha)} \nonumber
	\\ &\propto \Prob(\vA \given \vZ, \valpha) \Prob(\vZ \given 
	\valpha) \nonumber
	\\ &\propto \prod_{1 \le i < j \le n} p(Z_i, Z_j)^{a_{ij}} 
	\left(1 - p(Z_i, Z_j)\right)^{1 - a_{ij}} \prod_{k = 1}^{h} 
	Z_{ik}^{\alpha_k - 1}.
\end{align}
Since the density function expressed in 
Equation~(\ref{Eq:comlikelihood}) is not from any well-known 
distribution, we use another well-studied MCMC algorithm---the 
Metropolis Hastings sampling---to simulate the density function at 
each Gibbs iteration. We present the Gibbs sampling procedures in 
Algorithm~\ref{Alg:gibbs}. Notice that {\em burninNum} 
in the input of Algorithm~\ref{Alg:gibbs} refers to a burn-in 
number---a threshold of the Gibbs iterations, after which the 
distribution of our simulated samples converges to the posterior 
distribution of the target parameters. We thus only keep the 
simulated estimates after the burn-in number (as reflected in 
Line 11 in Algorithm~\ref{Alg:gibbs}). We usually choose a large 
burn-in number such that with a high probability, the MCMC iterations
have converged to the true posterior distribution.

\IncMargin{1em}
\begin{algorithm}[H]
	\KwIn{{\em burninNum} = 5000, {\em size} = 10000, empty set {\em 
			posteriorSample}}
	Initialization $Z_{ik} \leftarrow \frac{1}{h}$ for all $i = 1, 
	\ldots, n$ and $k = 1, \ldots, h$ \;
	Initialization {\em iterNum} $\leftarrow 1$ \; 
	\Repeat{iterNum $>$ size $+$ burninNum}{
		\For{$i = 1$ {\rm to} $n$}{
			Simulate $T_i \sim {\rm Dirichlet}(\valpha)$\;
			Simulate $U \sim {\rm Uniform}(0, 1)$\;
			\If{$U < \frac{\prod_{1 \le i \neq j \le n} p(T_i, 
					Z_j)^{a_{ij}} \left(1 - p(T_i, Z_j)\right)^{1 - 
						a_{ij}}\prod_{k = 1}^{h} T_{ik}^{\alpha_k - 
						1}}{\prod_{1 
						\le i \neq j \le n} p(Z_i, Z_j)^{a_{ij}} 
					\left(1 - 
					p(Z_i, Z_j)\right)^{1 - a_{ij}} \prod_{k = 
						1}^{h} 
					Z_{ik}^{\alpha_k - 1}}$}{
				Set $Z_i \leftarrow T_i$}}
		\If{iterNum $>$ burninNum}{Add $\vZ = (Z_1, Z_2, \ldots, 
			Z_n)$ to {\em posterioSample}}  
		Set {\em iterNum} $\leftarrow$ {\em iterNum}$+ 1$ \;
	}
	\KwOut{{\em posteriorSample}}
	\caption{The Gibbs sampling algorithm for the proposed mixed 
	membership model.}
	\label{Alg:gibbs}
\end{algorithm}
\DecMargin{1em}

After obtaining the {\em posteriorSample} of $\vZ$, we compute 
the sample mean $\bar{Z}_i$ as the Bayes estimate for $Z_i$, for 
each $i = 1, 2, \ldots, n$. For hard clustering, i.e., each of the 
nodes in the network only belongs to one community, so we assign 
every node to the community with the associated probability 
dominating the estimated membership parameter, i.e., $\argmax_{k} 
(\bar{Z}_{ik})$.

\section{Example: Zachary Karate Club Data}
\label{Sec:zachary}
In this section and the next, we evaluate the performance of our 
mixed membership model by applying it to two empirical social 
network data. The first that we consider is the Zachary karate 
club data, which was collected and used to study conflict and 
fission in small groups by~\citet{Zachary}. The data was from a 
university-based karate club of 34 members, who were tentatively 
divided into two groups due to an incipient conflict between the 
president of the club and the opposing faction. Consider the club as 
a social network consisting 34 nodes that represent club members. 
Each pair of the nodes are formalized by adding an edge in between 
if they are observed to interact outside normal activities, 
interpreted as ``extra'' friendship in~\citet{Zachary}. A total of 
78 (undirected) edges are observed; see~\citet[Figure 1]{Zachary}. 
The corresponding adjacency matrix was presented in~\citet[Figure 
2]{Zachary}.

\begin{table}[tbp]
	\centering
	\caption{Mixed membership result for the Zachary karate club 
		data.}
	\setlength{\tabcolsep}{8.6pt}
	\begin{tabular}{c c c c c c c c}
		\toprule
		Node & $Z_{i1}$ & $Z_{i2}$ & Cluster & Node & $Z_{i1}$ & 
		$Z_{i2}$ & Cluster \\ 
		\midrule
		$H$ & 0.8490 & 0.1510 & 1 & 18 & 0.7764 & 0.2236 & 1 \\  
		2 & 0.7190 & 0.2810 & 1 & 19 & 0.0683 & 0.9317 & 2 \\ 
		3 & 0.6144 & 0.3856 & 1 & 20 & 0.6932 & 0.3068 & 1 \\ 
		4 & 0.7034 & 0.2966 & 1 & 21 & 0.1063 & 0.8937 & 2 \\ 
		5 & 0.9466 & 0.0534 & 1 & 22 & 0.7694 & 0.2306 & 1 \\ 
		6 & 0.9741 & 0.0259 & 1 & 23 & 0.1181 & 0.8819 & 2 \\  
		7 & 0.9791 & 0.0209 & 1 & 24 & 0.2443 & 0.7557 & 2 \\ 
		8 & 0.6973 & 0.3027 & 1 & 25 & 0.3895 & 0.6105 & 2 \\ 
		9 & 0.1943 & 0.8057 & 2 & 26 & 0.3777 & 0.6223 & 2 \\ 
		10 & 0.5001 & 0.4999 & 1 & 27 & 0.1871 & 0.8129 & 2 \\ 
		11 & 0.9399 & 0.0601 & 1 & 28 & 0.3620 & 0.6380 & 2 \\ 
		12 & 0.7631 & 0.2369 & 1 & 29 & 0.4419 & 0.5581 & 2 \\ 
		13 & 0.7624 & 0.2376 & 1 & 30 & 0.1815 & 0.8185 & 2 \\ 
		14 & 0.6757 & 0.3243 & 1 & 31 & 0.1878 & 0.8122 & 2 \\ 
		15 & 0.0765 & 0.9235 & 2 & 32 & 0.3934 & 0.6066 & 2 \\ 
		16 & 0.0855 & 0.9145 & 2 & 33 & 0.0602 & 0.9398 & 2 \\ 
		17 & 0.9775 & 0.0225 & 1 & $A$ & 0.0880 & 0.9120 & 2 \\ 
		\bottomrule
	\end{tabular}
	\label{Table:zachary}
\end{table}

\begin{figure}[tbp]
	\centering
	\begin{subfigure}{0.5\textwidth}
		\centering
		\includegraphics[width=\linewidth]{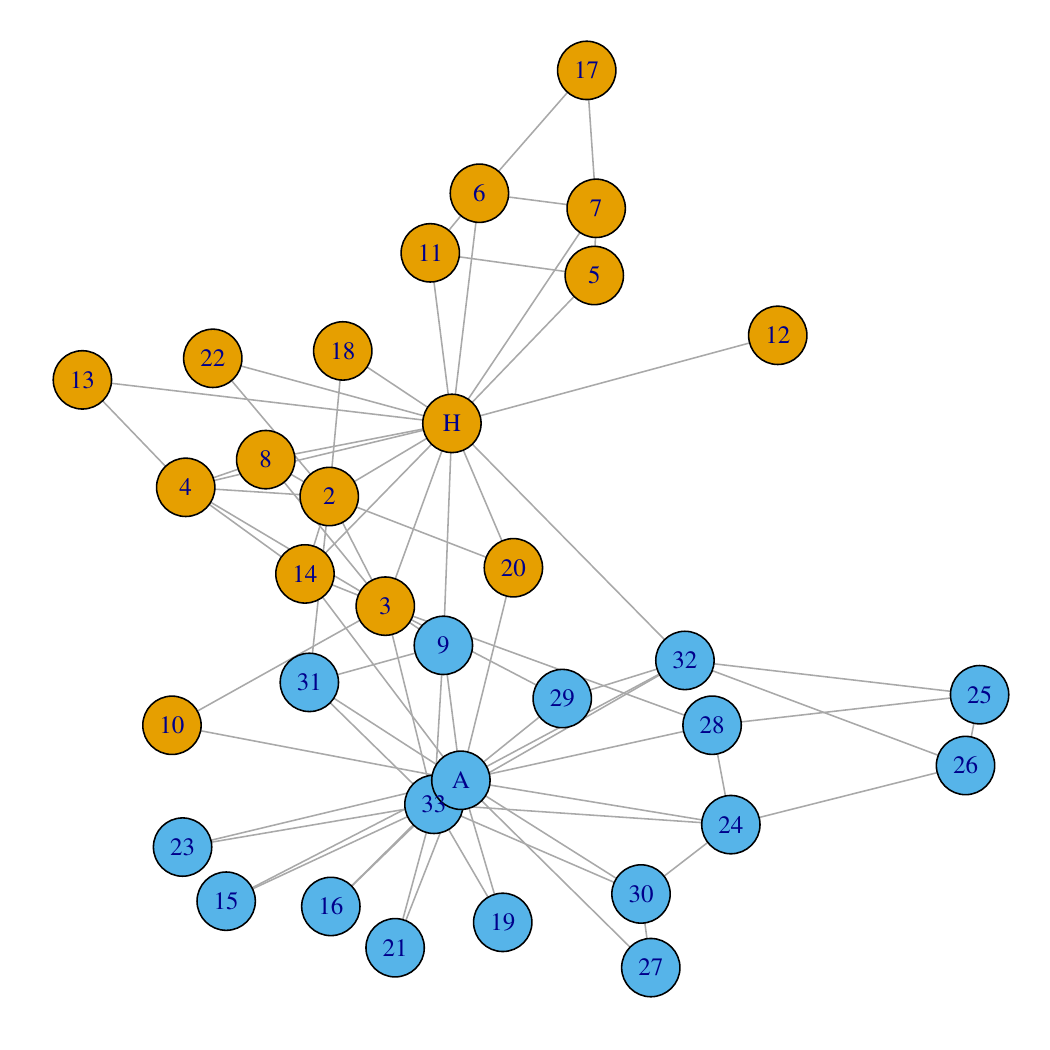}
		\caption{Clustering result of the Zachary karate club data 
		based on the proposed mixed membership model.} 
		\label{fig:karate_plot}
	\end{subfigure}%
	\begin{subfigure}{0.5\textwidth}
		\centering
		\includegraphics[width=\linewidth]{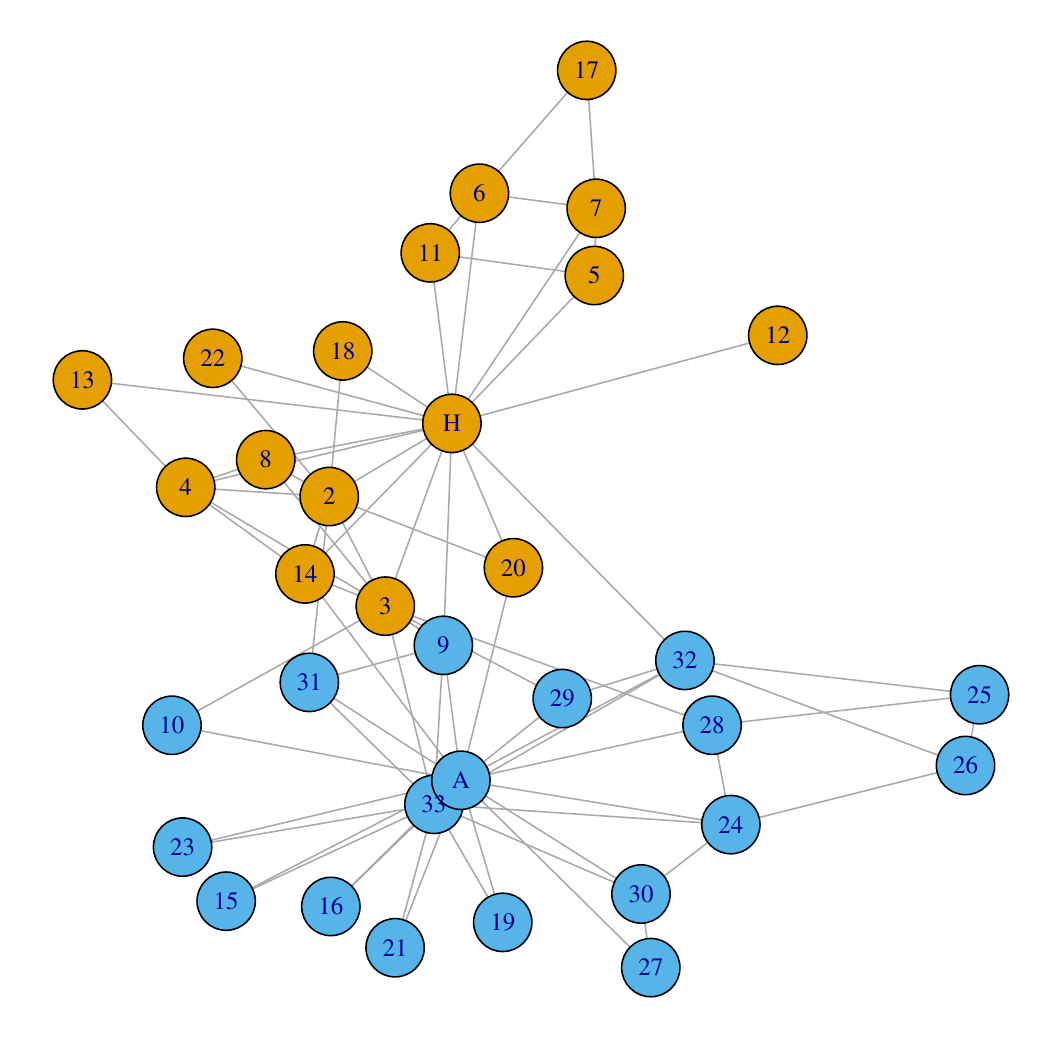}
		\caption{Ground truth of the clustering of the Zachary 
		karate club data.} \label{fig:karate_truth}
	\end{subfigure}
	\caption{Comparison between the clustering result and the ground 
	truth of the Zachary karate club data.} \label{fig:karate}
\end{figure}

We apply the mixed membership model proposed in 
Section~\ref{Sec:distance} to split the karate club members into 
two factions, and compare our clustering result with the ground 
truth released by~\citet{Zachary}. Based on the feature of the 
karate club network data and the background story, we set the number 
of communities $h = 2$. We implement Algorithm~\ref{Alg:gibbs}, 
for which the {\em burninNum} and {\em size} respectively take 
values 5,000 and 10,000. The posterior mean $\bar{Z}_{ik}$, for 
$i = 1, 2, \ldots, n$ and $k = 1, 2$, is used as the Bayes estimate 
for the mixed membership parameter $Z_{ik}$. The result is presented 
in Table~\ref{Table:zachary}. If a hard clustering framework is 
considered, we present a graphic summary in 
Figure~\ref{fig:karate_plot}, where the nodes in different 
clusters are distinguished by different colors: orange for 
community $1$ and blue for community $2$.

For the purpose of comparison, the ground truth corresponding 
to~\citet[Figure 1]{Zachary} and~\citet[Table 1]{Zachary} is 
portrayed in Figure~\ref{fig:karate_truth}. We observe that the 
entity labeled with $10$ is the only misclassified node according to 
our model. We cluster node $10$ into community $1$, but in reality 
node $10$ joins community $2$. The occurrence of misclassification 
of node $10$ is probably because the node is connected with one node 
(node $3$) from community $1$, and is also connected with one node 
(node $A$) from community $2$. However, node $A$ is the center of 
community $2$, hence more influential in the network. Additionally, 
Table~\ref{Table:zachary} shows that the membership 
parameter estimate for node $10$ is $0.5001$ for community $1$ 
versus $0.4999$ for community $2$, so the difference is minimal.

\section{Example: Bottlenose Dolphin Network Data}
\label{Sec:dolphin}
In this section, we analyze the bottlenose dolphin network data 
from~\citet{Lusseau}. A study of identifying the roles that 
bottlenose dolphins played in their social network was conducted 
by~\citet{Lusseau2004}. The network data was collected for 62 
bottlenose dolphins living in Doubtful Sound, New Zealand, over 
a period of seven years from 1994 to 2001. The bottlenose dolphins 
are represented by nodes in the network, and ties between nodes are 
interpreted as associations between dolphin pairs occurring more 
often (due to some sort of homophily) than expected by chance. 
There is a total of 318 edges observed in the network. 

\begin{table}[tbp]
	\centering
	\caption{Mixed membership result for the bottlenose dolphin 
		network data.}
	\setlength{\tabcolsep}{8.6pt}
	\begin{tabular}{c c c c c c c c}
		\toprule
		Node & $Z_{i1}$ & $Z_{i2}$ & Cluster & Node & $Z_{i1}$ & 
		$Z_{i2}$ & Cluster \\ 
		\midrule
		1 & 0.0126 & 0.9874 & 2 & 32 & 0.3661 & 0.6339 & 2 \\ 
		2 & 0.1125 & 0.8875 & 2 & 33 & 0.3610 & 0.6390 & 2 \\ 
		3 & 0.0029 & 0.9971 & 2 & 34 & 0.7588 & 0.2412 & 1 \\ 
		4 & 0.9005 & 0.0995 & 1 & 35 & 0.6634 & 0.3366 & 1 \\ 
		5 & 0.5303 & 0.4697 & 1 & 36 & 0.5429 & 0.4571 & 1 \\ 
		6 & 0.2785 & 0.7215 & 2 & 37 & 0.8460 & 0.1540 & 1 \\ 
		7 & 0.2313 & 0.7687 & 2 & 38 & 0.8132 & 0.1868 & 1 \\ 
		8 & 0.0721 & 0.9279 & 2 & 39 & 0.7190 & 0.2810 & 1 \\ 
		9 & 0.8991 & 0.1009 & 1 & 40 & 0.4481 & 0.5519 & 2 \\ 
		10 & 0.2386 & 0.7614 & 2 & 41 & 0.8124 & 0.1876 & 1 \\ 
		11 & 0.0082 & 0.9918 & 2 & 42 & 0.1883 & 0.8117 & 2 \\ 
		12 & 0.5440 & 0.4560 & 1 & 43 & 0.0080 & 0.9920 & 2 \\ 
		13 & 0.5295 & 0.4705 & 1 & 44 & 0.7278 & 0.2722 & 1 \\ 
		14 & 0.2291 & 0.7709 & 2 & 45 & 0.6549 & 0.3451 & 1 \\ 
		15 & 0.7747 & 0.2253 & 1 & 46 & 0.9868 & 0.0132 & 1 \\ 
		16 & 0.9554 & 0.0446 & 1 & 47 & 0.5691 & 0.4309 & 1 \\ 
		17 & 0.7469 & 0.2531 & 1 & 48 & 0.0175 & 0.9825 & 2 \\ 
		18 & 0.2039 & 0.7961 & 2 & 49 & 0.3791 & 0.6209 & 2 \\ 
		19 & 0.9943 & 0.0057 & 1 & 50 & 0.5713 & 0.4287 & 1 \\ 
		20 & 0.0717 & 0.9283 & 2 & 51 & 0.7497 & 0.2503 & 1 \\ 
		21 & 0.7326 & 0.2674 & 1 & 52 & 0.9920 & 0.0080 & 1 \\ 
		22 & 0.9909 & 0.0091 & 1 & 53 & 0.7884 & 0.2116 & 1 \\ 
		23 & 0.3629 & 0.6371 & 2 & 54 & 0.5116 & 0.4884 & 1 \\ 
		24 & 0.9187 & 0.0813 & 1 & 55 & 0.1423 & 0.8577 & 2 \\ 
		25 & 0.9905 & 0.0095 & 1 & 56 & 0.9420 & 0.0580 & 1 \\ 
		26 & 0.1431 & 0.8569 & 2 & 57 & 0.3022 & 0.6978 & 2 \\ 
		27 & 0.1348 & 0.8652 & 2 & 58 & 0.2325 & 0.7675 & 2 \\ 
		28 & 0.1317 & 0.8683 & 2 & 59 & 0.5491 & 0.4509 & 1 \\ 
		29 & 0.0403 & 0.9597 & 2 & 60 & 0.9095 & 0.0905 & 1 \\ 
		30 & 0.9912 & 0.0088 & 1 & 61 & 0.3816 & 0.6184 & 2 \\ 
		31 & 0.0400 & 0.9600 & 2 & 62 & 0.5052 & 0.4948 & 1 \\
		\bottomrule
	\end{tabular}
	\label{Table:dolphins}
\end{table}

\begin{figure}[tbp]
	\centering
	\begin{subfigure}{0.5\textwidth}
		\centering
		\includegraphics[width=\linewidth]{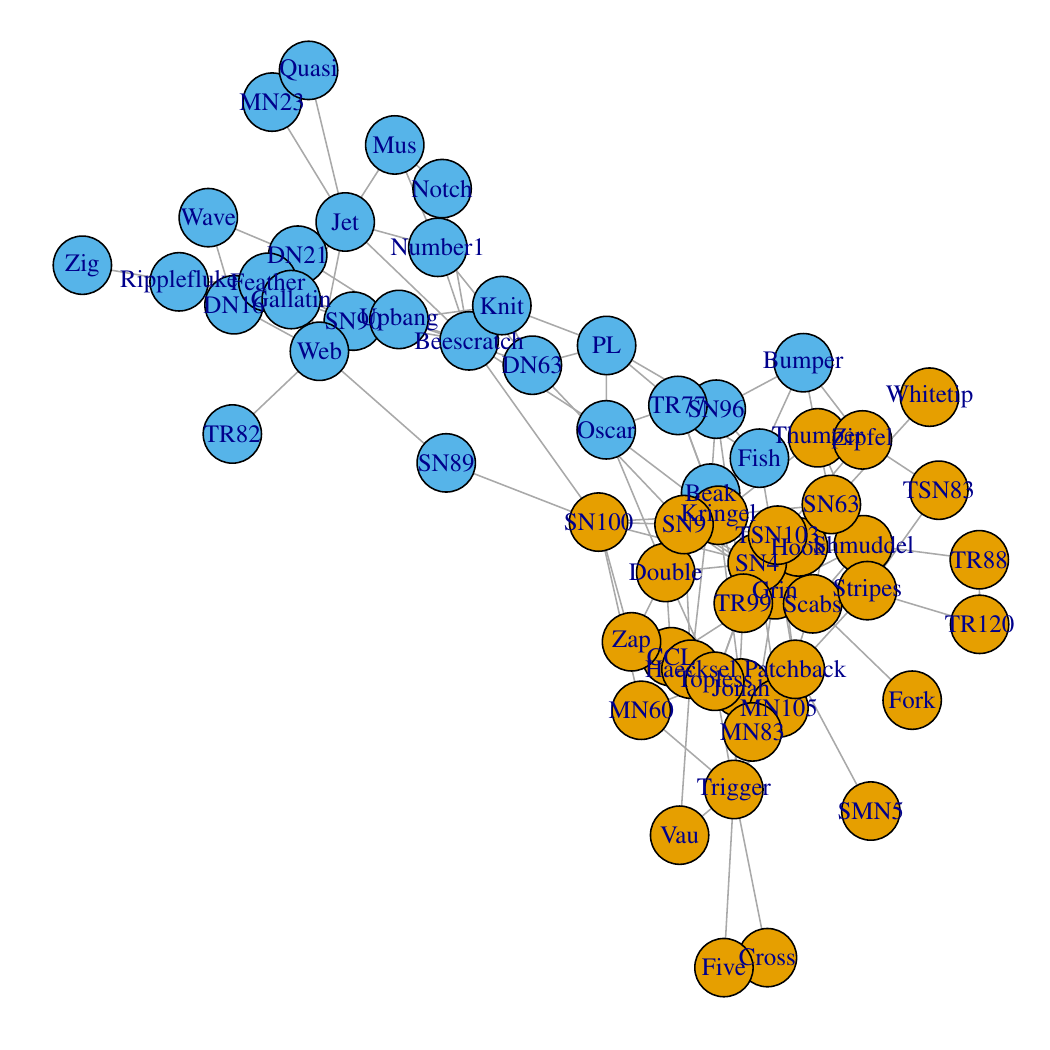}
		\caption{Clustering result of the bottlenose dolphin network 
		data based on the proposed model.} 
		\label{fig:dolphins_plot}
	\end{subfigure}%
	\begin{subfigure}{0.5\textwidth}
		\centering
		\includegraphics[width=\linewidth]{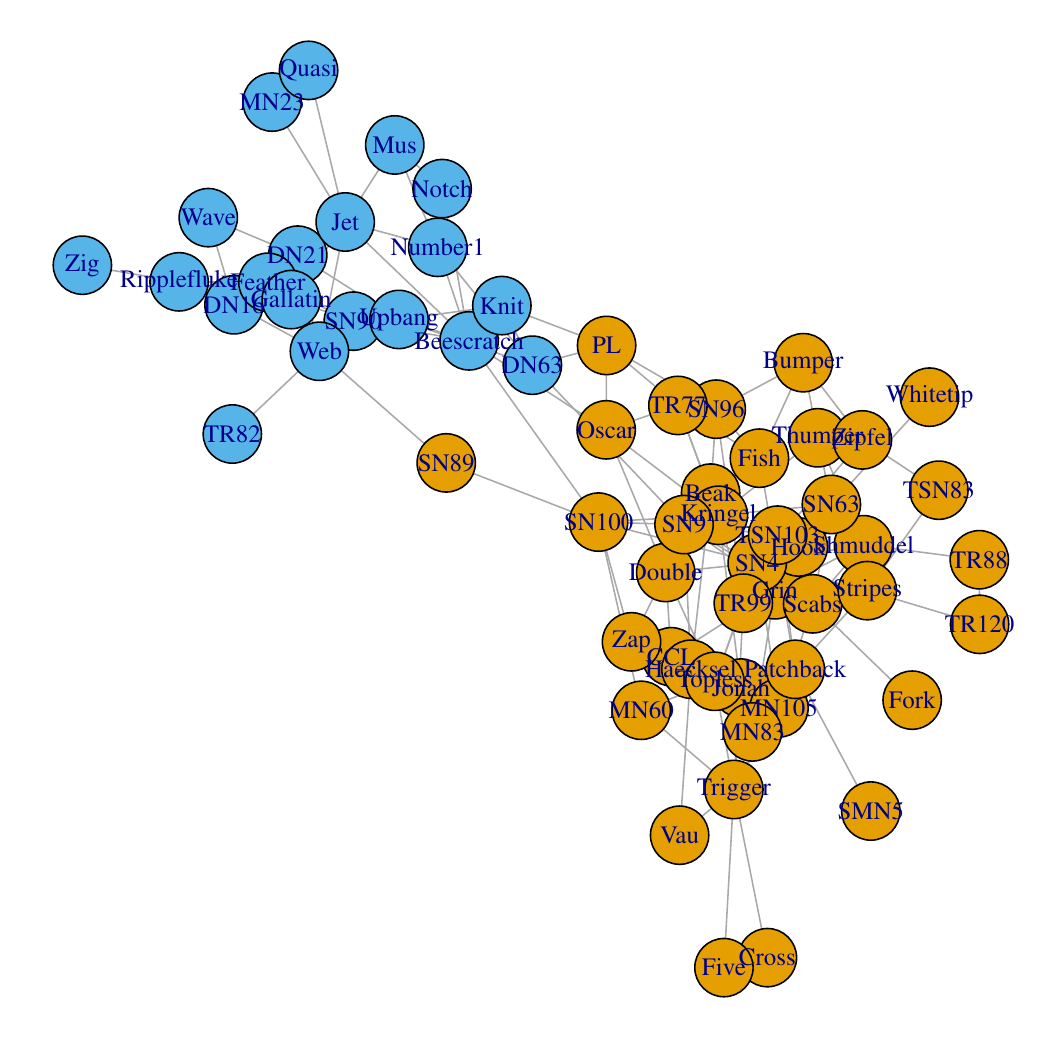}
		\caption{Ground truth of the clustering of the bottlenose 
		dolphin network data.} \label{fig:dolphins_truth}
	\end{subfigure}
	\caption{Comparison between the clustering result and the ground 
		truth of the bottlenose dolphin network data.} 
		\label{fig:dolphins}
\end{figure}

A natural division of the bottlenose dolphin network was 
discussed in~\citet{Lusseau2004}, and it was done via an accurate 
and sensitive clustering algorithm proposed by~\citet{Girvan}. The 
algorithm therein was based on a newly-defined ``betweenness'' 
measure generalized from the one defined in~\citet{Freeman}. Two 
communities were detected for the bottlenose dolphin network, shown 
in~\citet[Figure 1(a)]{Lusseau2004}, as well as in 
Figure~\ref{fig:dolphins_truth}, for the purpose of comparison. 

We set $h = 2$ in our mixed membership model based on the 
conclusion from~\citet{Lusseau2004}. Both of the {\em burninNum} and 
{\em size} take value 50000. Executing Algorithm~\ref{Alg:gibbs} 
with the new {\em burninNum} and {\em size}, we obtain the mixed 
membership probabilities for the bottlenose dolphins, organized in 
Table~\ref{Table:dolphins}. We also depict the hard clustering 
result in Figure~\ref{Fig:bayesiandolphin} for a better 
visualization. The nodes in community $1$ are colored 
with orange, while the nodes in community $2$ are colored with blue.

Comparing the clustering result of our mixed membership model 
and that of the betweenness-based model in~\citet{Lusseau2004}, we 
realize that the community classification matches for most of 
the nodes in the network, except for ``Beak'', ``Bumper'', 
``Fish'',  ``Oscar'', ``PL'', ``SN89'', ``SN96'' and ``TR77'' on the 
boundary. \citet{Lusseau2004}, in fact, assigned these dolphins to 
a sub-community of Community $1$ using their algorithm.

\section{Simulations}
\label{Sec:simulation}
We show the identifiability and reliability of our model as well 
as the proposed algorithm through two empirical social network 
examples in Sections~\ref{Sec:zachary} and~\ref{Sec:dolphin}. 
However, both of those networks only contain a relatively small 
number of nodes which are only divided into two communities. In this 
section, we run a few more simulations to further evaluate the 
performance of our algorithm. We simulate several SBMs with 
different predetermined community structure. Each block in the 
simulated SBMs is generated by implementing an algorithm for the 
\ER\ graph~\citep{Gilbert}. There are three key parameters for 
simulated SBMs: class size, within-cluster link density and 
cross-cluster link density. Noticing that the community structure of the 
simulated networks is known, we can use this information as the 
ground truth for assessment.

Two well-defined metrics, the Normalized Mutual Information 
({\rm NMI})~\citep{Meila} and the Adjusted Rand Index ({\rm 
ARI})~\citep{Rand}, are adopted to examine the closeness 
between the clustering results of our algorithm and the ground 
truths. In addition, we implement another commonly-used method for 
social network clustering---the modularity maximization 
algorithm~\citep{Newman02}---to the simulated SBMs for further 
comparison.

We simulate a total of five SBMs with known community structure as 
summarized in Table~\ref{tab:sbm}. {\rm SBM1} and {\rm SBM2} are 
both in moderate size (i.e., of order 100), containing two 
communities (of sizes 80 and 20, respectively). The 
within-cluster link densities are significantly high (0.8) for 
both SBMs, whereas the cross-cluster link density of {\rm SBM1} is 
much smaller than that of {\rm SBM2} (0.02 v.s.\ 0.2). Next, we 
consider a larger network. We simulate {\rm SBM3} of order 
500, of which 400 nodes form one community, and the rest 100 nodes 
form the other. Although one community is much lager than the other, 
both of the community sizes are generally large in {\rm SBM3}. Then, 
we consider networks containing more than two communities. There are 
three communities in {\rm SBM4} 
and four communities in {\rm SBM5}, respectively. In {\rm SBM4}, the 
size of one community (100) is significantly larger than those 
of the other two (10 for each). Empirically, extremely small-size 
communities are likely to cause problems for network clustering. 
In {\rm SBM5}, all the communities are quite close in size.
\begin{table}[tbp]
	\centering
	\caption{Link density summaries for the simulated SBMs.} 
	\label{tab:sbm}
	\setlength{\tabcolsep}{6.3pt}
	\begin{tabular}{c cc cc cc ccc cccc}		
		\toprule
		& \multicolumn{2}{c}{SBM1} & \multicolumn{2}{c}{SBM2} & 
		\multicolumn{2}{c}{SBM3} & \multicolumn{3}{c}{SBM4} & 
		\multicolumn{4}{c}{SBM5}
		\\
		\cmidrule(lr){2-3} \cmidrule(lr){4-5} \cmidrule(lr){6-7} 
		\cmidrule(lr){8-10} \cmidrule(lr){11-14}
		& $C_1$ & $C_2$ & $C_1$ & $C_2$ & $C_1$ & $C_2$ & $C_1$ & 
		$C_2$ & $C_3$ & $C_1$ & $C_2$ & $C_3$ & $C_4$ \\
		\midrule
		size & 80 & 20 & 80 & 20 & 400 & 100 & 100 & 10 & 10 & 40 & 
		30 & 20 & 20 \\
		\midrule
		$C_1$ & $0.8$ & $0.02$ & $0.8$ & $0.2$ & $0.8$ & $0.02$ & 
		$0.8$ & $0.02$ & $0.02$ & $0.8$ & $0.02$ & $0.02$ & $0.02$ 
		\\
		$C_2$ & $0.02$ & $0.8$ & $0.2$ & $0.8$ & $0.02$ & $0.8$ & 
		$0.02$ & $0.8$ & $0.02$ & $0.02$ & $0.8$ & $0.02$ & $0.02$
		\\
		$C_3$ & & & & & & &$0.02$ & $0.02$ & $0.8$ & $0.02$ & $0.02$ 
		& $0.8$ & $0.02$
		\\
		$C_4$ & & & & & & & & & & $0.02$ & $0.02$ & $0.02$ & $0.8$
		\\
		\bottomrule
	\end{tabular}
\end{table}

For each SBM, we set {\em burninNum} at 1000 and {\em size} at 
2000, respectively. The proposed algorithm is run for 30 times, 
and for each result, both {\rm ARI} and {\rm NMI} are computed. The 
averages of all 30 {\rm ARI}'s (i.e., $\widehat{\rm ARI}$) and 
{\rm NMI}'s (i.e., $\widehat{\rm NMI}$) are used as estimates for 
evaluating the performance of the algorithm. In addition, we 
implement the modularity maximization algorithm to all five 
simulated SBMs, and compute the corresponding {\rm ARI} and {\rm 
NMI}. These results are presented in Table~\ref{tab:res}.
\begin{table}[ht]
	\centering
	\caption{Evaluation of clustering results.} 
	\label{tab:res}
	\setlength{\tabcolsep}{8.9pt}
	\begin{tabular}{c c c c c}
		\toprule
		& \multicolumn{2}{c}{Algorithm~\ref{Alg:gibbs}} & 
		\multicolumn{2}{c}{Mod. max.}
		\\ \cmidrule(lr){2-3} \cmidrule(lr){4-5}
		& $\widehat{\rm ARI}$ & $\widehat{\rm NMI}$ & {\rm ARI} 
		& {\rm NMI}
		\\ \midrule
		{\rm SBM1} & 0.8215 & 0.7392 & 0.2661 & 0.4102 
		\\
		{\rm SBM2} & 0.7246 & 0.6466 & 0.2025 & 0.2555
		\\
		{\rm SBM3} & 0.8863 & 0.8112 & 1.0000 & 1.0000
		\\
		{\rm SBM4} & 0.8772 & 0.7633 & 0.2194 & 0.3218
		\\
		{\rm SBM5} & 1.0000 & 1.0000 & 1.0000 & 1.0000
		\\
		\bottomrule
	\end{tabular}
\end{table}

We observe that the proposed algorithm performs well in general 
for all simulated SBMs. On the other hand, it seems that the 
modularity maximization algorithm undergoes several severe 
clustering problems. The first problem that we notice is 
over-clustering. In theory, there is no cluster structure in the 
\ER\ graph. However, the modularity maximization algorithm divides 
predetermined communities (i.e., the \ER\ graphs) to reach a higher 
modularity index for small networks, reflected in the clustering 
results for {\rm SBM1} and {\rm SBM2}. Both of the clustering 
results indicate that four communities are needed for these two 
simulated networks so as to attain the global maximum of the 
modularity index. Second, the modularity maximization algorithm also 
has under-clustering problem sometimes, especially when communities 
are extremely small. In {\rm SBM4}, the modularity index reaches the 
global maximum when the two smaller communities merge together. The 
inconsistency of the modularity maximization algorithm was discussed 
extensively by~\citet{Bickle}. However, it seems that the modularity 
maximization algorithm overperforms when all the communities are 
large in size, for instance, {\rm SBM3}. Besides, our algorithm and 
the modularity maximization algorithm both perform perfectly well 
when the sizes of communities are similar in the network. 
Nevertheless, we conclude that the proposed algorithm is more robust 
for social network clustering.

\section{Concluding Remarks}
\label{Sec:concluding}

In this paper, we develop a simple but novel model-based method 
for social network clustering. We adopt the cosine function to 
measure similarities between nodes. In addition, we propose an 
algorithm based on the Gibbs sampling to simulate posterior samples 
for mixed community membership for entities in the network. Our 
model is not only flexible for fuzzy clustering, but also amenable 
for hard clustering. We would like to point out that our model is 
reliable due to solid theoretical foundation of Bayesian approach 
and MCMC algorithms. We evaluate the performance of our model 
through two empirical social network data and simulations. Based on 
comparisons with ground truth, we conclude that our model provides 
accurate clustering for social network data

At last, we discuss several limitations of our model, and propose 
some future studies. First, it is known that MCMC algorithms are 
slow to achieve stationary distribution. The complexity of the 
proposed algorithm in this paper is $O(n^2)$ for each Gibbs 
iteration. In addition, a large number of {\em burninNum} is usually 
needed for ensuring convergence. Admittedly, the algorithm is not 
efficient especially when the number of parameters or the size of 
network data or both are large. There is an urge of developing 
faster algorithms for our mixed membership model. One alternative is 
the Hamiltonian Monte Carlo (HMC) algorithm, which can 
accelerate convergence to the target distribution by simulating 
Hamiltonian dynamics. We refer the interested readers 
to~\citet{Neal} for a detailed explanation of HMC, and 
to~\citet{Betancourt} for an exposition of the intuition 
behind HMC. Another possible approach is to use variational Bayesian 
methods to convert simulation procedures to optimization 
problems, and then implement some appropriate approximation 
algorithms.

Second, our current model itself can be improved. (1) 
The proposed model measures node relationship based on cosine 
similarity, which is analogous to Pearson correlation, so it may 
fail to preserve membership homophily for the nodes close or on the 
cluster boundary. These nodes usually have similar entries in their 
corresponding membership variables, but the proposed model favors 
connections among these nodes regardless of their actual membership 
information. For instance, suppose $Z_i = (0.51, 0.49)$ and $Z_j = 
(0.49, 0.51)$, we then have $p(Z_i, Z_j) = 0.999$ albeit $i$ and 
$j$ belonging to different communities in our setting; (2) The 
proposed model does not focus on sparse networks particularly. One 
may consider tweaking cosine similarity as $p(Z_i, Z_j) = \rho_n 
\cos(Z_i, Z_j)$ with a scaling factor $\rho_n \to 0$ as $n$ 
increases to incorporate network sparsity. 
the proposed model yet accounts for the information possibly 
contained in the nodes. We would like to consider a more complete 
model which utilizes those auxiliary variables so as to further 
improve clustering accuracy.

Third, the communities considered in 
our model are distinct. One of our future work is to look into a 
possibility to extend our model to 
overlapping communities like in~\citet{Xie}.

Lastly, our model, 
as well as most other graphical generative models, requires a prior 
knowledge about the number of communities to which nodes are 
assigned. However, this number is usually unavailable. Estimating 
the number of communities and membership parameters simultaneously 
could be a challenging task. A recent research paper~\citep{Geng} 
provides us some guidance about future study in this direction.

\end{document}